\documentclass[final,preprint,aps,showpacs,fleqn]{revtex4}
\usepackage{bm}
\usepackage[dvipsnames]{xcolor} 
\usepackage{graphicx}
\usepackage{setspace}
\usepackage{amsmath,amsfonts,amssymb,amscd,amsxtra,amsthm}
\usepackage{graphicx}  
\usepackage{epstopdf}
\usepackage{color}
\usepackage{dcolumn}  
\usepackage{bm}          
\usepackage{multirow}
\usepackage[utf8]{inputenc}
\usepackage[normalem]{ulem} 

\newcommand{\bea}{\begin{eqnarray}}
\newcommand{\eea}{\end{eqnarray}}
\newcommand{\re}[1]{(\ref{#1})}
\newcommand{\tr}{{\rm tr}}

\begin{document}




\title{Dynamical gluon mass in the instanton vacuum model}

\author{M. Musakhanov}
\email{musakhanov@gmail.com}
\author{	O. Egamberdiev}
\affiliation{Theoretical Physics Department, National University of
  Uzbekistan, Tashkent 
100174, Uzbekistan}

\begin{abstract}
We   {consider} the modifications of gluon
properties in  {the} instanton liquid model (ILM) for the QCD vacuum. 
Rescattering of  gluons on  instantons  {generates} the dynamical momentum-dependent gluon mass
$M_g(q).$  First, we   {consider the case of a
  scalar gluon},  no zero-mode problem
 {occurs} and  its dynamical mass $M_s(q)$  {can be
  found}.    {Using} the typical phenomenological values
of the average instanton size $\rho=1/3\,\,fm$ and average
inter-instanton distance $R=1\,\, fm$ we   {get}
$M_s(0)=256\,\,MeV$.  We  {then} extend this
approach to the real vector gluon with  zero-modes  {carefully considered}.   {We obtain the following
  expression } $M^2_g(q)=2M^2_s(q).$ This modification of the gluon in
the instanton media  {will shed light on
  nonperturbative aspect on heavy quarkonium physics}. 
\end{abstract}
\maketitle


\section{Introduction}
Without any doubt instantons represent a very important topologically
nontrivial component of the QCD vacuum. The main parameters of the QCD
instanton vacuum  { developed in the instanton liquid model (ILM)}  
are the average instanton size $\rho$ and inter-instanton distance $R$
(see,  {for example},  {following
  reviews}~\cite{Diakonov:2002fq,Schafer:1996wv}).  
They were phenomenologically estimated as $\rho = 1/3\, \mathrm{fm},\,
R = 1\, \mathrm{fm}$ and confirmed by theoretical variational
calculations~\cite{Diakonov:2002fq,Schafer:1996wv}  
and recent lattice simulations of the QCD vacuum
 \cite{Chu:1993fc,Negele:1998ev,DeGrand:2001tm,Faccioli:2003qz,Bowman:2004xi}. 
 In particular, the spontaneous breakdown of chiral symmetry is realized very
well via the ILM~\cite{Goeke:2007bj}. Hence, instantons play a pivotal and significant
role in describing the lightest hadrons and their interactions. 

In  {the} ILM  {the} instanton induces  strong
interactions between light quarks   {and}
  {produce} a large dynamical mass $M$, 
 {which was initially almost massless.}   {Consequently, light quarks are bound and the
pions as a pseudo-Goldstone boson appear as a result of spontaneous 
breakdown of chiral symmetry (SBCS).}
  
On the other hand,   {the
  instantons from the QCD vacuum} also interact with heavy quarks and
 {are} responsible for the generation of the heavy-heavy and
heavy-light quark interactions with trace of the
SBCS~\cite{Musakhanov:2011xx,Musakhanov:2014fya,Musakhanov:2017gym}. It
is important   {to note that the packing parameter is
  given as $\rho^4/R^4$ and it become very small ($\sim 0.01$) with
the phenomenological values of $\rho$ and $R$ used.}
 {Then, the dynamical quark mass $M$ is expressed as
  $(packing\,\,parameter)^{1/2}\rho^{-1}\sim 365$
  MeV~\cite{Diakonov:2002fq} while the instanton contribution to the
  heavy quark mass is given as $\Delta M\sim
(packing\,\,parameter)\rho^{-1}\sim 70$ MeV~\cite{Diakonov:1989un}}. 
We see that these specific $packing\,\,parameter$ dependencies explain
the values of $M$ and $\Delta M$. These factors define the coupling
between  {the} light-light, heavy-light and heavy-heavy quarks
induced by   {the instantons from the
  QCD vacuum}. 

  {The direct
instanton effects  mainly contribute to the intermediate 
region characterized by the instanton size ($\rho\simeq 0.33$ fm), as
was studied in Ref.~\cite{Turimov:2016adx} in which the instanton
effects are marginal but still important to be considered for a
quantitative description of the heavy quarkonium spectra. One-gluon exchange
is dominant at smaller distances.} 
On the other hand,  {a size of the heavy quarkonium is}
small~\cite{Digal:2005ht}. It means that heavy quarkonium properties
might be sensitive to a modification of gluon properties in
instanton media (ILM) induced by rescattering of 
 {gluons} on   {the} instantons.  

Previously the dynamical gluon mass $M_g$ within ILM was estimated
in~\cite{Hutter:1995sc} as $M_g\sim 400\,MeV$   {with}
phenomenological values of $\rho$ and $R$. 
 {However,} this estimation was 
 {obtained by}  ignoring  {the} gluon zero-modes problem and
some $SU(N_c)$ factors.   

  {In this work, we aim at
investigating the dynamical gluon mass within the ILM, extending the
method developed in Ref.~\cite{Pobylitsa:1989uq}, where the formulae
for the quark correlators were derive.}   

\section{Scalar "gluon" propagator}
We start from the scalar massless field $\phi$ belonging to the
adjoint representation as a real gluon. We have to find its propagator
in the external classical gluon field   {in the ILM} 
$A_\mu=\sum_I A^I_\mu (\gamma_I)$, where $ A^I_\mu (\gamma_I)$ is a
generic notation for the QCD (anti-) instanton in the singular gauge.
 {$\gamma_I$ stands for all the relevant
collective coordinates: the position in
Euclid 4D space $z_I$, the size $\rho_I$ and the $SU(N_c)$ color
orientation $U_I$. The number of the collective coordinates is $4N_c$.} 

\begin{figure}[h]
\centerline{\includegraphics[scale=0.8]{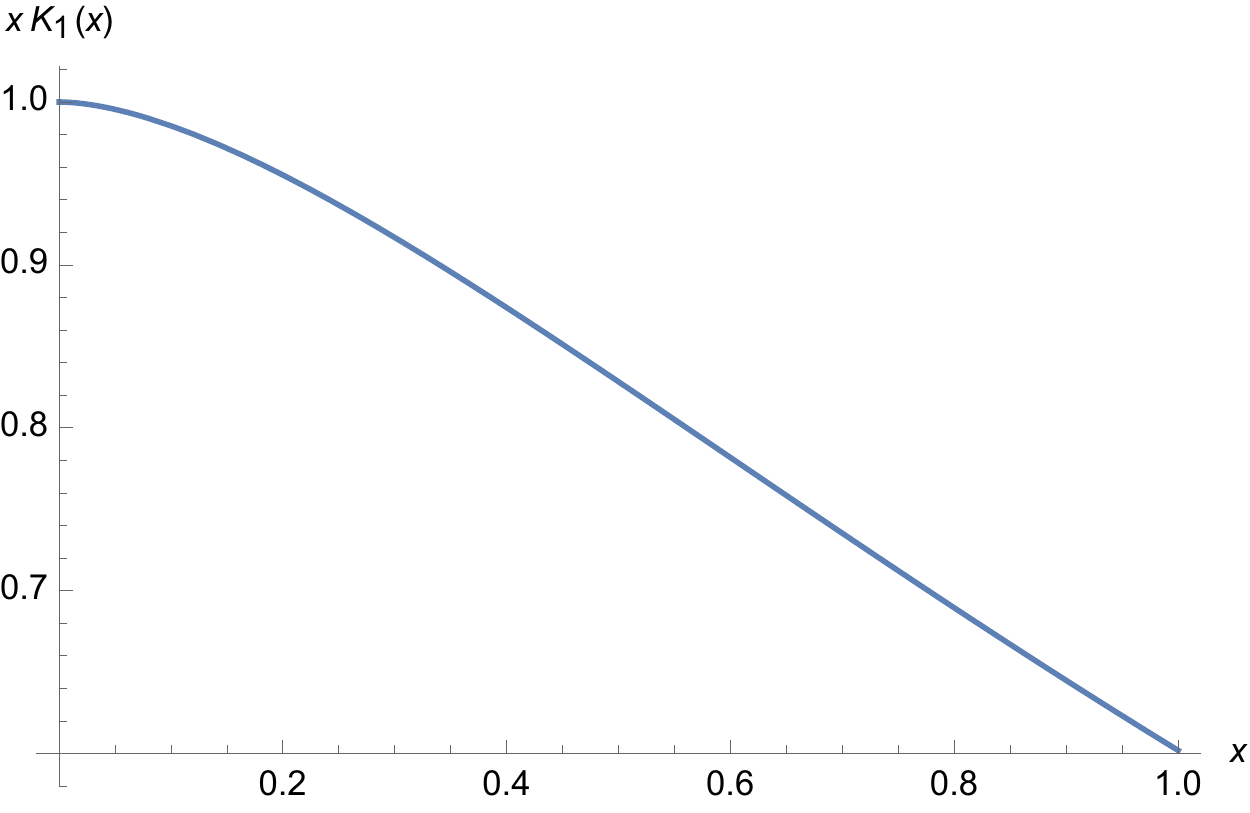}}
\caption{Scalar "gluon" dynamical mass form-factor as a function of $x=q\rho$.}
\label{K1}
\end{figure}

The action is defined as $S_\phi=(\phi^+ P^2\phi) $ where $P_\mu=p_\mu
+ A_\mu$ (in the coordinate representation $p_\mu=i\partial_\mu$). The
scalar gluon-like propagator is given by 
\bea
&&\Delta=(p+A)^{-2}=(p^2+\sum_i(\{p,A_i\}+A_i^2)+\sum_{i\neq
  j}A_iA_j)^{-1},\,\,\,\Delta_0=p^{-2}, 
\\\nonumber
&&\tilde{\Delta}=(p^2+\sum_i(\{p,A_i\}+A_i^2))^{-1},\, \,\,\,
\Delta_i=P_i^{-2}=(p^2+\{p,A_i\}+A_i^2)^{-1}. 
\eea
There  {are} no zero modes in $\Delta^{-1}_i=P_i^2$ and
$\Delta^{-1}=P^2$, 
  {which} means  {the} existence
of the inverse operators $\Delta_i$ and  $\Delta$. Our aim is to find
the propagator averaged over instanton collective coordinates  
$
\bar\Delta\equiv<\Delta>=\int D\gamma \,\Delta .
$
  {However,} we start first from
$
\bar{\tilde{\Delta}}.
$
 Expanding $\tilde\Delta$ over $(\{p,A_i\}+A_i^2)$
  {carrying out further resummation}, 
 {we obtain} the multi-scattering series 
\bea
\tilde\Delta=\Delta_0 +\sum_i (\Delta_i-\Delta_0)+  \sum_{i\neq
  j}(\Delta_i-\Delta_0)\Delta_0^{-1}(\Delta_j-\Delta_0) 
+...
\eea 
  {As in Ref.}~\cite{Pobylitsa:1989uq}, the main
contribution to the $\bar{\tilde\Delta}$ can be 
 {summed up} by  {the following equation } 
\bea
\bar{\tilde\Delta}-\Delta_0=\sum_i<\{{\bar{\tilde\Delta}}^{-1}(\Delta_0^{-1}-\Delta_i^{-1})^{-1}\Delta_i^{-1}-(\Delta_0^{-1}-{\bar{\tilde\Delta}}^{-1})\}^{-1}> 
\label{PobylEq1}\eea
Rewriting this equation, we have
\bea
\bar{\tilde\Delta}^{-1}-\Delta_0^{-1}=\sum_i<\{\bar{\tilde\Delta}
+ (\Delta_i^{-1}-\Delta_0^{-1})^{-1}\}^{-1}>
\label{PobylEq2}\eea
  {We can derive} the solution of
 Eqs.~\re{PobylEq1} and \re{PobylEq2} in  {the} ILM by expanding
  {with respect to the} instanton density $N/V=1/R^4$,
since the actual dimensionless expansion parameter  is
 {in fact} the packing parameter  {$\rho^4/R^4$}. 
The solution of   {Eq.}~\re{PobylEq2} 
 {in} the  {first-order expansion with respect to the} 
density   {comes from} the iteration if  {replaces the right-hand side of this
equation by}  
$\bar{\tilde\Delta}\rightarrow\Delta_0$. Then we have 
\bea
\bar{\tilde\Delta}^{-1}-\Delta_0^{-1}&=&<\sum_i\{\Delta_0
+ (\Delta_i^{-1}-\Delta_0^{-1})^{-1}\}^{-1}>
\nonumber\\
&=&-<\sum_i\Delta_0^{-1}(\Delta_i-\Delta_0)\Delta_0^{-1}>= N\Delta_0^{-1}(\bar\Delta_I-\Delta_0)\Delta_0^{-1},
\label{Eqdelta}\eea
where $\bar\Delta_I=\int d\gamma_I\, \Delta_I $.
Now  {we} compare $\Delta$   {with}
$\tilde\Delta$.   {Expanding} $\Delta$
  {with respect to} $A_iA_j$,   {we get}
\bea
\Delta=\tilde\Delta - \tilde\Delta \sum_{i\neq j}A_iA_j\Delta= 
 \tilde\Delta - \tilde\Delta \sum_{i\neq j}A_iA_j\tilde\Delta +...
\label{Deltaexpansion}\eea
It means immediately 
$\bar{\Delta}-\bar{\tilde\Delta}=O(N^2)$ which is negligible. 
 We have to take  {a} well-known result{s} for 
$\Delta_I $ from  {Ref.}~\cite{Brown:1977eb}: 
\bea
&&\Delta^{ab}_I= \frac{1}{2}\tr\frac{\tau_a F(x,y)\tau_b F(y,x)}{4\pi^2(x-y)^2 \Pi(x)\Pi(y)},\,\,\,\,\Pi(x)=\frac{x^2+\rho^2}{x^2} ,
\label{Delta}\\
&&\tau_\mu=(\vec\tau,i),\,\,\,\tau^+_\mu =(\vec\tau,-i),\,\,\,
\tau_\mu\tau^+_\nu=\delta_{\mu\nu}+i\bar\eta_{a\mu\nu}\tau_a,
\\
&& F(x,y)=1+\rho^2\frac{(\tau x)( \tau^+ y)}{x^2y^2}=1+\rho^2\frac{(xy)}{x^2y^2}+\rho^2\frac{i\bar\eta_{a\mu\nu}\tau_a x_\mu y_\nu}{x^2y^2},
\eea
  {where} $\bar\eta_{a\mu\nu}=-\bar\eta_{a\nu\mu}$ is
the    {'tHooft}   {symbol.}  
  {We assume that}  {the} position of the
instanton $z=0$ and the orientation $U=1$. It is clear  {to see}  
from  Eq.~\re{Eqdelta} that the  gluon-like scalar
dynamical mass operator is given by 
\bea
M_s^2\delta_{ab}=<\sum_i p^2(\Delta^{ab}_i-\Delta^{ab}_0)p^2>=
N (p^2\bar\Delta^{ab}_Ip^2-\delta_{ab}p^2).
\eea
 {In order to average over the position $z$,}  we have to change
$x\rightarrow x-z$, $y\rightarrow y-z$ and 
 {perform integration} $\int d^4z$.   {Similarly, we average over the color orientation $U$.} 
  {Introducing the} orientation  factor
$O^{ab}=\tr  (U^+t^aU\tau^b)$,  {where} $t_a$ are $SU(N_c)$- matrices, we
change $\Delta^{ab}_I$  {to be} $O^{ab}O^{a'b'}\Delta^{bb'}_I$, and
  {carry out integration}  $\int dO$. 
Here $\int dO O^{ab}O^{ab'}=\delta_{bb'},  \int dO O^{ab}O^{a'b'}=(N_c^2-1)^{-1}\delta_{aa'}\delta_{bb'}$. 
Also, $\int dO O^{ab}\bar\eta_{b\mu\nu}O^{a'b'}\bar\eta_{b'\mu'\nu'}=(N_c^2-1)^{-1}\delta_{aa'}(\delta_{\mu\mu'}\delta_{\nu\nu'}-\delta_{\mu\nu'}\delta_{\nu\mu'}). $
In coordinate space,  {we find} 
\bea
 \nonumber
&&\bar\Delta^{aa'}_I (x,y)-\Delta^{aa'}_0 (x,y)
 \\ \nonumber
 &&=\int d^4z dO  O^{ac}O^{a'c'}(\Delta^{cc'}_I (x',y')-\Delta^{cc'}_0 (x',y'))\,\,\,(x'\equiv x-z,\,\,\,y'\equiv y-z),
\\
&&=\delta_{aa'}\int d^4z [\frac{3\rho^2}{4\pi^2(N_c^2-1)} f_1(x')f_1(y') + \frac{2\rho^4}{N_c^2-1}f_{2}(x')g(x'-y')f_{2}(y')],
\label{barDelta}\\ \nonumber
&&f_1(x)=\frac{1}{(x^2+\rho^2)},\,\,\,f_2(x)= \frac{(x_\mu x_\nu, ix^2)}{x^2(x^2+\rho^2)},\,\,\,\,
g(x-y)=\frac{1}{4\pi^2(x-y)^2}.
\eea
  {In momentum space, we}
find the contribution from the first term in  {Eq.}~\re{barDelta} as 
 \bea 
 M_{1,s}(q)=[\frac{3\rho^2}{(N_c^2-1)R^4} 4\pi^2]^{1/2} q\rho K_1(q\rho)
\label{M1}
\eea 
where the   {form factor} $q\rho
K_1(q\rho)$. $K_1$ denotes the  {modified} Bessel function.  {In Fig.~\ref{K1}, we draw the form factor.}
The  {estimation} of the contribution from the
second term in  Eq.~\re{barDelta} leads to $ M_{2,s}(0)=0$,
while   {in general} we know  $
M_{2,s}(q\rightarrow \infty)\rightarrow 0.$ It means  {that} we 
may neglect this contribution at all and  {obtain} $ M_{s}(q)= M_{1,s}(q)$. 
 {Using the phenomenological values of $\rho$ and $R$, we obtain
  $M_s(0)=256\,MeV$.} 
\section{Real gluon propagator}
 {The}   {total} gluon field in the ILM is $A+a$,
where $A=\sum_i A^i(\gamma_i)$.   {The} number
of all collective coordinates $\gamma_i$  {is} equal to $4N_c
N$. We have to  {decompose}   {the
so-called zero modes $\phi_\mu^{i}$ from the total 
fluctuation $a$,} which are the fluctuations 
 along the collective coordinates $\gamma_i$ in the 
functional space.  Consider first  {the} single instanton case,
  {based on
  Refs.}~\cite{Vandoren:2008xg,Brown:1978bta}. All of the fluctuations
will be taken with gauge fixing condition $P^I_\mu a_\mu=0$
 {imposed}. Then, the quadratic part of
the effective action is  {given as} $(a_\mu M^I_{\mu\nu}a_\nu)$,
where $M^I_{\mu\nu}={P^I}^2\delta_{\mu\nu}+2iG^I_{\mu\nu}-(1-1/\xi)P^I_\mu
P^I_\nu $ and $G^I_{\mu\nu}=-i[P^I_\mu,P^I_\nu]$. 
Here  $\xi$   {stands for the} gauge fixing parameter.
The zero modes are the solutions of    {the following equation} 
 \bea
M^I_{\mu\nu}\phi^{i}_\nu=0.
\eea
In some sense  {the} zero modes   {can be considered
  as}  derivatives   {with respect to} collective coordinates 
of the instanton field together with the additional longitudinal term
dictated by  {the} gauge fixing condition. 
The   {projection operator} to the instanton
zero-modes space is  {defined as} $P^I_{\mu\nu}=\sum_i
\phi^{i}_\mu\phi^{i+}_\nu$, while  {that}  to the
nonzero modes space is  {defined as}
$Q^I_{\mu\nu}=\delta_{\mu\nu}-P^I_{\mu\nu}.$  
The gluon propagator $S^I_{\mu\nu}$ is defined by 
 {the following equation} 
\bea
M^I_{\mu\nu}S^I_{\nu\rho}=Q^I_{\mu\rho}
\eea
The explicit solution of this   {equation}  was
 {already} derived in  {Ref.}~\cite{Brown:1977eb}.
  {To generalize the
  formulae in Ref.~\cite{Pobylitsa:1989uq},} we introduce  {an} artificial gluon mass $m$,which will be
taken zero  {at the end of calculation}. 
So, we define $g^I_{m,\mu\nu}$ and    {take the limit
  of} $\lim_{m\rightarrow 0}
g^I_{m,\mu\nu}=S^I_{\mu\nu}, $
where $$(M^I_{\mu\rho}+m^2\delta_{\mu\rho})
g^I_{m,\rho\nu}=Q^I_{\mu\nu}. $$ 
  {We also} introduce $G^I_{m,\rho\nu}$,   {satisfying}
$$(M^I_{\mu\rho}+m^2\delta_{\mu\rho}) G^I_{m,\rho\nu}=\delta_{\mu\nu}. $$
As  {was} shown in  {Ref.}~\cite{Brown:1978bta},  {we find}
$$ G^I_{m,\rho\nu}=g^I_{m,\rho\nu}+\frac{1}{m^2}P^I_{\rho\nu}.$$
It is clear  {to see} that 
$$ {G^I}^{-1}_{m,\mu\rho}=(M^I_{\mu\rho}+m^2\delta_{\mu\rho}).$$
Now we may repeat the  {the same method with} which
  {we are able to obtain} the averaged ILM "scalar"
gluon propagator $\bar{\tilde\Delta}$ given   {in}
Eqs. (\ref{PobylEq1},\ref{PobylEq2}). First,  {we} introduce
 {the} ILM inverse massive gluon propagator 
 $$ {G}^{-1}_{m,\mu\rho}={P}^2\delta_{\mu\nu}+2iG_{\mu\nu}+m^2\delta_{\mu\rho}= \tilde M_{\mu\nu}+m^2\delta_{\mu\rho} +\sum_{i\neq j}(A^iA^j\delta_{\mu\nu} -i[ A^i_\mu,A^j_\nu]).  $$
and 
$$\tilde G^{-1}_{m,\mu\nu}=\tilde M_{\mu\nu}+m^2\delta_{\mu\nu}= 
p^2+\sum_i ((\{p,A^i\}+{A^i}^2)\delta_{\mu\nu}+2iG^i_{\mu\nu})+m^2\delta_{\mu\nu}.$$
  {Following the way we have derived}
Eqs. (\ref{PobylEq1},\ref{PobylEq2}) and neglecting $O(\rho^8/R^8)$
terms, we  can  {immediately} find the averaged   
$\bar{\tilde G}_{m,\mu\nu}$   {as follows}
\bea
&&\bar{\tilde G}_{m,\rho\nu}-G^0_{m,\rho\nu}=\sum_i<\{{{ G^0}^{-1}_{m,\rho\mu}}({G^0}_{m,\mu\alpha}^{-1}
-{G^i_{m,\mu\alpha}}^{-1})^{-1}{G^i_{m,\alpha\nu}}^{-1}\}^{-1}>
\nonumber\\\label{barG}
&&=\sum_i<(G^i_{m,\alpha\nu}- {G^0}_{m,\rho\nu})>=N(\bar G^I_{m,\alpha\nu}- {G^0}_{m,\rho\nu} )
\eea
and   {equivalently}
\bea
\bar{\tilde G}_{m,\rho\nu}-G^0_{m,\rho\nu}=N(\bar S^I_{m,\rho\nu}-{S^0}_{m,\mu\nu} +\frac{1}{m^2}\bar P^I_{\rho\nu})
\eea
We finally see $\bar{\tilde G}_{m,\rho\nu} =\bar{S}_{m,\rho\nu}+
\frac{1}{m^2}\bar P_{\rho\nu} $, where  $\bar P_{\rho\nu}=N\bar P^I_{\rho\nu}$.
 {The} ILM non-zero modes propagator $\bar{S}_{m,\rho\nu}$
  {in the limit of} $m\rightarrow 0$ limit become $ \bar{
  S}_{\rho\nu}$ and  is given by 
\bea
\bar{ S}_{\rho\nu}-S^0_{\rho\nu}=N(\bar S^I_{\rho\nu}-{S^0}_{\rho\nu}),
\label{barS}\eea
where ${S^0}_{\mu\nu}=(\delta_{\mu\nu}-(1-\xi)p_\mu p_\nu/p^2)/p^2$ is
the free gluon propagator. It is obvious  {to see} that 
${S^0}^{-1}_{\mu\nu}=\delta_{\mu\nu}p^2-(1-1/\xi)p_\mu p_\nu$.

We expect $ \bar{ S}_{\rho\nu}=(\delta_{\mu\nu}-(1-\xi)p_\mu
p_\nu/p^2)/(p^2+M_g^2) .$   {Thus,
  we are able} to rewrite Eq.\re{barS} in another equivalent form
\bea
M_g^2\delta_{\rho\nu}=N{S^0}^{-1}_{\rho\sigma}(\bar S^I_{\sigma\mu}-{S^0}_{\sigma\mu}){S^0}^{-1}_{\mu\alpha}(\delta_{\alpha\nu}-(1-\xi)p_\alpha p_\nu/p^2).
\label{Mg}\eea
This   {equation}  {defines the}
gauge-invariant ($\xi$-independent) dynamical gluon mass $M_g.$ 

Here  {the} single instanton gluon propagator \cite{Brown:1977eb}
is  {given as}
\bea
S^I_{\mu\nu}=q_{\mu\nu\rho\sigma}P^I_\rho\Delta_I^2 P^I_\sigma -(1-\xi)P^I_\mu \Delta_I^2 P^I_\nu,
\label{SI}
\eea 
where
$q^I_{\mu\nu\rho\sigma}=\delta_{\mu\nu}\delta_{\rho\sigma}+\delta_{\mu\rho}\delta_{\nu\sigma}
-\delta_{\mu\sigma}\delta_{\nu\rho}+ \epsilon_{\mu\nu \rho\sigma}$.

From  Eq. \re{barDelta} we know that in  coordinate space
   {the most slowly decreasing part}  part of the $\Delta_I$  {in the limit of
$x\rightarrow\infty$ and similarly $y\rightarrow\infty$}  is given by the first
term ($\sim f_1(x-z)f_1(y-z)$) of  {Eq.}~\re{barDelta}. Only this term gave a
contribution to the $M_s$.  {A} similar analysis  can be
  {done} for   $M_g$. 
We expect that   {the most slowly decreasing part} part
$S^I_{\nu\mu}-{S^0}_{\nu\mu}$ will  {only contribute to}
 $M_g.$ 
In  coordinate space  {we find}
$P^I_\mu=i\partial_\mu+A_\mu^I(x-z)$ and
$A^I_\mu(x)=\bar\eta_{a\mu\nu}\tau_a\frac{x_\nu\rho^2}{x^2(x^2+\rho^2)}$. 
Comparing  {the} effects  from $i\partial_\mu$ 
  {with} $A^I_\mu$, we conclude from  {Eq.}~\re{SI}
that the   {the most slowly decreasing part} part of the
$S^I_{\nu\mu}-{S^0}_{\nu\mu}$ in Eq. \re{Mg}   {comes} from 
$$
p_\rho(  {the\,\, most\,\, slowly\,\, decreasing } \,\,part\,\,of\,\,(\Delta_I-\Delta_0)\Delta_0+ \Delta_0(\Delta_I-\Delta_0)) p_\sigma 
$$   
and only this term will   {contribute} to
  $M_g$. Comparing  {it} with  Eq. \re{M1}, we
conclude that $M^2_g(q)=2M^2_s(q), $ where $q$ dependence is
represented by Fig.\ref{K1}.   {Using} the
phenomenological values of $\rho$ and $R$,  {we obtain}
$M_g(0)=362\,MeV$. 
\section{Conclusion}
The strength of the gluon-instanton interaction is given by dynamical
gluon mass $M_g$ and  {it} is large. It   {depends} on the parameters
$\rho$ and $R$  {as in the case of the dynamical light quark mass}: 

 \centerline{ $M_g\sim (packing\,\, parameter)^{1/2}\,\,\rho^{-1}\sim 362\,MeV.$ }

  {This modification of the gluon propagator from
  the instanton vacuum will provide the Yukawa-type potential in
  addition to all other pieces with instanton effects recently reported in
  Ref.~\cite{Turimov:2016adx}.} Since the 
 {distance}  {that is} most sensitive to this modification
  {is approximately around} $r_g\approx
M_g^{-1}= 0.55\,fm$, we conclude that   {it may still give some effects on charmonium
properties.}   {Further investigation is under way.}  

 {\bf Acknowledgement.} M.M.  {is} thankful to Hyun-Chul Kim for the useful
 and helpful communications. This work is supported by Uz grant
 OT-F2-10. 

\section*{References}
\bibliography{gluonmass310118PLB}
\end{document}